\documentstyle[iopconf1,epsfig]{article}
\begin{document}

\def\simlt{\stackrel{<}{{}_\sim}}
\def\simgt{\stackrel{>}{{}_\sim}}

\vspace{-4 in}
\rightline{
KAIST-TH 97/17}
\rightline{FTUAM 97/20}

\title{Charge and Color Breaking in 
Supersymmetry and Superstring\footnote{Based on talks given at
`Beyond the desert: accelerator and non--accelerator approaches',
Castle Ringberg, Tegernsee (Germany), June 1997;
`8th Miniworkshop on Particle and Astroparticle Physics',
Pusan (South Korea), May 1997.}}

\author{Carlos Mu\~noz\dag\footnote{Permanent address:
Departamento de F\'{\i}sica Te\'orica, Universidad Aut\'onoma de
Madrid, Cantoblanco, 28049 Madrid, Spain.}} 

\affil{\dag\ Department of Physics, Korea Advanced Institute of Science
and Technology, Taejon 305-701, Korea}


\beginabstract
Charge and color 
breaking minima in SUSY theories might 
make the standard vacuum unstable.
In this talk a brief review of this issue is performed.
When a complete analysis of all the potentially dangerous directions in the
field space of the theory is carried out, 
imposing that the standard vacuum should be the global minimum, 
the corresponding constraints
turn out to be very strong and, in fact, there are extensive regions in the
parameter space of soft SUSY--breaking terms that become forbidden.
For instance, in the context of the MSSM
with universal soft terms,
this produces important bounds, not only on the value of $A$, but also
on the values of $B$, $M$ and $m$. 
In specific SUSY scenarios, as fixed point models, 
no--scale supergravity, 
gauge--mediated SUSY breaking and superstrings, the
charge and color breaking constraints are also very important.
For example, 
if the dilaton is the source of SUSY breaking
in four--dimensional superstrings,
the whole parameter space ($m_{3/2}$,$B$) 
is excluded on these grounds. Cosmological analyses are also
briefly reviewed.
\endabstract

\section{Introduction and summary}
As is well known, 
the presence of scalar fields with color and electric charge in
supersymmetric (SUSY) theories induces the possible existence of
dangerous charge and color breaking minima, which would make
the standard vacuum unstable \cite{reciente}. 
This is not necessarily a shortcoming since many SUSY models can
be discarded on these grounds, thus improving the predictive power
of the theory. 
A complete
analysis of all the potentially dangerous directions in the field
space of the minimal supersymmetric standard model (MSSM) was carried
out in \cite{CCB1}. It was shown there that,
imposing that the SUSY standard vacuum should be {\it deeper} than
the charge and color breaking minima, the corresponding constraints
on the soft parameter space 
are very strong (see also \cite{olive}). 
For instance, in the universal case and assuming
radiative symmetry breaking with nothing but the MSSM in between the
weak scale and the grand unification scale $M_{GUT}$,
there are extensive regions of this space that
become {\it forbidden} producing
important bounds, not only on the value of the trilinear scalar parameter 
($A$), but also on the values of the bilinear scalar parameter ($B$) and the
scalar and gaugino masses ($m, M$ respectively). 
The above mentioned constraints were used 
in \cite{baer} finding a
preferred region of SUSY particle masses after imposing in addition
dark matter and naturalness constraints. 
Very strong bounds can also be obtained applying the above mentioned 
constraints to particular
SUSY scenarios. This is the case of the infrared fixed 
point model \cite{CCB2} and a SO(10) GUT \cite{strumia}.
No--scale supergravity models where the limit $m=0$ is obtained would be
excluded on these grounds \cite{CCB1,CCB2}.
Charge and color breaking constraints were also studied in the context
of gauge--mediated SUSY--breaking models \cite{randall}. In most of them
the global vacuum does not preserve QCD. 
On the other hand, 
the stability of the corresponding
constraints with respect to variations of the initial scale for the
running of the soft breaking parameters
was analyzed in \cite{CCB2}, finding that the larger the scale
is, the stronger the bounds become. In particular, by taking the Planck
scale rather than $M_{GUT}$ for the initial scale, substantially stronger
constraints are found.
These issues are
reviewed in section 2. Let us finally remark that the stability of the
standard vacuum also imposes constraints on flavor--mixing trilinear
soft terms which are stronger than the laboratory bounds coming from
the absence of FCNC \cite{dimo}.

The low--energy limit of four--dimensional superstrings is a 
SUSY field theory. This allows us to apply the above mentioned general
constraints to SUSY/string scenarios. The analysis can be in principle
more
predictive since in four--dimensional superstrings 
it is possible to obtain
information about the structure of soft SUSY--breaking terms
\cite{BIM3}. 
In particular,
in the dilaton--dominated SUSY--breaking scenario, 
the soft terms 
are universal and depend on only two parameters, the gravitino
mass $m_{3/2}$ and $B$. 
It was shown in \cite{CCB3} that 
charge and color breaking
constraints are so important that
the {\it whole} parameter space is forbidden and, as a consequence, the
dilaton--dominated SUSY breaking is excluded on these grounds.
In section 3 this analysis is reviewed. The possibility
of assuming that the moduli fields contribute to SUSY breaking is
also discussed.

Finally, section 4 is left for some final comments including 
cosmological considerations.

\section{Charge and color breaking in supersymmetry}

A complete study of this crucial issue is in principle very involved. This
is mainly due to two reasons. First, the enormous complexity of the
scalar potential, $V$, in a SUSY theory. Second, the radiative corrections
to $V$ must be included in a proper way. Concerning the first point,
the tree--level scalar potential, using a standard notation, is given by
$V_0 = V_F + V_D + V_{\rm soft}$, with
%
%
\begin{eqnarray}
\label{VF}
V_F = \sum_\alpha \left| \frac{\partial W}{\partial \phi_\alpha}
\right| ^2\;\;,\; \;
V_D = \frac{1}{2}\sum_a g_a^2\left(\sum_\alpha\phi_\alpha^\dagger
T^a \phi_\alpha\right)^2\;\; ,
\end{eqnarray}
\begin{eqnarray}
\label{Vsoft}
V_{\rm soft}&=&\sum_\alpha m_{\alpha}^2
|\phi_\alpha|^2\ +\ \sum_{i\equiv generations}\left\{
A_{u_i}\lambda_{u_i}Q_i H_2 u_i + A_{d_i}\lambda_{d_i} Q_i H_1 d_i
\right.
\nonumber \\
&+& \left. A_{e_i}\lambda_{e_i}L_i H_1 {e_i} + {\rm h.c.} \right\}
+ \left( B\mu H_1 H_2 + {\rm h.c.}\right)\;\; ,
\end{eqnarray}
where $W$ is the MSSM superpotential
\begin{eqnarray}
\label{W}
W=\sum_{i\equiv generations}\left\{
\lambda_{u_i}Q_i H_2 u_i + \lambda_{d_i}Q_i H_1 d_i
+ \lambda_{e_i} L_i H_1 e_i \right\} +  \mu H_1 H_2\;\; ,
\end{eqnarray}
and $\alpha$ runs over all the canonically normalized 
scalar components of the chiral
superfields. 
The first observation is that the 
previous potential is extremely
involved since it  has a large number of independent fields.
Furthermore, even assuming universality of the soft breaking terms at
$M_{GUT}$, it contains a large number of
independent parameters: $m$, $M$, $A$, $B$, $\mu$.
In addition, there are the
gauge ($g$)
and Yukawa ($\lambda$)
couplings which are constrained by the experimental data. Notice that
$M$ does not appear explicitely in $V_0$, but it does through the
renormalization group equations (RGEs) of all the remaining parameters.

Concerning the radiative corrections it should be noted that
the tree--level scalar potential $V_0$ is strongly dependent on
the renormalization scale $Q$, and the one--loop radiative corrections
to it, namely
$\Delta V_1=\sum_{\alpha}\frac{n_\alpha}{64\pi^2}
M_\alpha^4\left[\log\frac{M_\alpha^2}{Q^2}-\frac{3}{2}\right]$,
%
%
are crucial to make the potential stable against variations of the $Q$ scale.
In the previous expression 
$M_\alpha^2(Q)$ are the improved tree--level
squared mass eigenstates and $n_\alpha=(-1)^{2s_\alpha} (2s_\alpha+1)$, 
where $s_\alpha$ is the spin of the corresponding particle. 
Clearly the complete one--loop potential $V_1=V_0+\Delta V_1$
has a structure that is even far more involved than $V_0$. Notice 
that $M_\alpha^2(Q)$ are in general
field--dependent quantities since they are the eigenvalues of
the $(\partial^2 V_0/\partial \phi_i\partial\phi_j)$ matrix. Hence, the 
values of $M_\alpha^2(Q)$ depend on the values of the fields and thus
on which direction in the field space is being analyzed.
This makes in practice the minimization of the complete $V_1$ an
impossible task. However, in the region of $Q$ where $\Delta V_1$ 
is small, the predictions of $V_0$ and $V_1$ essentially coincide.
This occurs for a value of $Q$ of the order of the most significant 
$M_{\alpha}$ mass appearing in 
$\Delta V_1$, which in turns
depends on what is the direction in the field space that is being
analyzed. Therefore one can still work just with $V_0$, but with the
approximate choice of $Q$. 

Taking into account all the above points one should carry out a complete
analysis of all the possible dangerous directions in the field space
along which the potential develops a charge and color breaking
minimum deeper than the realistic one.
The latter, given by
$V_{\rm real\;min}
=- \frac{1}{8} (g'^2+g_2^2) (v_2^2-v_1^2)^2$, 
%
%
where $|H_1^0|=v_1$, $|H_2^0|=v_2$ with 
$v_1^2+v_2^2=2M_W^2 / g_2^2$, corresponds to the standard vacuum.
Several comments with respect to this minimum are in order. 
First, note that result $V_{\rm real\;min}$ is obtained by minimizing
just the tree-level part of the Higgs potential.
As explained above this procedure is correct if the minimization
is performed at some sensible scale $Q\equiv M_S$, which should be of the order
of the most relevant mass entering $\Delta V_1$.
Since we are dealing here with the 
Higgs--dependent part of the potential, that mass is necessarily
of the order of the largest Higgs--dependent mass, namely the
largest stop mass. A more precise estimate of $M_S$, using a certain average
of typical SUSY masses, can be found in 
\cite{CCB1}. 
Second,
the requirement of correct electroweak breaking fixes one
of the five independent parameters of the MSSM, say $\mu$, in terms
of the others ($m$,$M$,$A$,$B$). Third, we must be sure that the realistic
minimum is really a minimum in the whole field space. This simply implies
that all the scalar squared mass eigenvalues (charged Higgses, squarks
and sleptons) must be positive. This is guaranteed for the charged Higgs
fields since in the MSSM the minimum of the Higgs potential always lies at
$H_2^+=H_1^-=0$, but not for the rest of the sparticles. Finally, we must
go further and demand that all the not yet observed particles
have masses compatible with the
experimental bounds. 
%

There are two types of charge and color breaking constraints:
the ones arising from directions in the field--space along
which the (tree--level) potential can become unbounded from below (UFB),
and those arising from the existence of charge and color
breaking (CCB) minima in the potential deeper than the
standard minimum. Since it is not possible to give here an account of
the explicit form of the constraints we refer the interested reader
to 
\cite{CCB1}. Here we will mention only their most 
important characteristics.

Concerning the CCB constraints, let us mention that the 
``traditional'' 
bound, first studied by Frere et al. and subsequently by
others \cite{reciente}, when correctly evaluated (i.e. including the
radiative corrections in a proper way) turns out to be extremely weak.
However, the 
``improved'' 
set of analytic constraints obtained in
\cite{CCB1}, which represent the necessary and sufficient conditions
to avoid dangerous CCB minima, is much stronger.

Concerning the UFB directions (and corresponding constraints),
there are three of them, labelled as UFB--1, UFB--2, UFB--3
in \cite{CCB1}. It is worth mentioning here that in general the
unboundedness is only true
at tree level since radiative corrections eventually raise the potential for
large enough values of the fields, but still these minima can be deeper than
the realistic one and thus dangerous.
The UFB--3 direction, which involves
the fields
$\{H_2,\nu_{L_i},e_{L_j},e_{R_j}\}$ with $i \neq j$
and thus leads also to electric charge
breaking, yields the {\it strongest} bound among {\it all}
the UFB and CCB constraints so it deserves to be exposed in greater detail. 
The explicit form of this bound
is as follows.
By simple analytical minimization it is possible to write the
value of all the relevant fields along the UFB--3 direction in
terms of the $H_2$ one. Then, for any value of $|H_2|<M_{GUT}$ satisfying
\begin{eqnarray}
\label{SU6}
|H_2| > \sqrt{ \frac{\mu^2}{4\lambda_{e_j}^2}
+ \frac{4m_{L_i}^2}{g'^2+g_2^2}}-\frac{|\mu|}{2\lambda_{e_j}} \ ,
\end{eqnarray}
the value of the potential along the UFB-3 direction is simply given
by
\begin{eqnarray}
\label{SU8}
V_{\rm UFB-3}&=&(m_2^2 -\mu^2+ m_{L_i}^2 )|H_2|^2
+ \frac{|\mu|}{\lambda_{e_j}} ( m_{L_j}^2+m_{e_j}^2+m_{L_i}^2 ) |H_2|
\nonumber\\ &&
-\ \frac{2m_{L_i}^4}{g'^2+g_2^2} \ .
\end{eqnarray}
Otherwise
\begin{eqnarray}
\label{SU9}
V_{\rm UFB-3}&=& (m_2^2 -\mu^2 ) |H_2|^2
+ \frac{|\mu|} {\lambda_{e_j}} ( m_{L_j}^2+m_{e_j}^2 ) |H_2| 
\nonumber\\ &&
+\ \frac{1}{8}
(g'^2+g_2^2)\left[ |H_2|^2+\frac{|\mu|}{\lambda_{e_j}}|H_2|\right]^2 \ .
\end{eqnarray}
In (\ref{SU8}) and (\ref{SU9}) $\lambda_{e_j}$ is the leptonic Yukawa
coupling of the $j-$generation and $m_2^2$ is the sum of the $H_2$ squared
soft mass, $m_{H_2}^2$, plus $\mu^2$. Then, the
UFB--3 condition reads
$V_{\rm UFB-3}(Q=\hat Q) > V_{\rm real \; min}$,
%
%
where $V_{\rm real \; min}$
was given above 
and the $\hat Q$ scale is given by 
$\hat Q\sim {\rm Max}(g_2 |e|, \lambda_{top} |H_2|,
g_2 |H_2|, g_2 |L_i|, M_S)$
with
$|e|$=$\sqrt{\frac{|\mu|}{\lambda_{e_j}}|H_2|}$ and
$|L_i|^2$=$-\frac{4m_{L_i}^2}{g'^2+g_2^2}$ 
+($|H_2|^2$+$|e|^2$).
Notice from (\ref{SU8}) and (\ref{SU9}) 
that the negative contribution to $V_{UFB-3}$
is essentially given by the $m_2^2-\mu^2$ term, which can be very sizeable in 
many instances. On the other hand, the positive contribution is dominated by 
the term $\propto 1/\lambda_{e_j}$, thus the larger
$\lambda_{e_j}$ the more restrictive
the constraint becomes. Consequently, the optimum choice of
the $e$--type slepton is the third generation one, i.e.
${e_j}=$ stau.

Now, we will analyze numerically the above constraints. We will see that
they are very important and, in fact, there are {\it extensive regions}
in the parameter space which {\it are forbidden}. 
Our analysis will be quite general in the sense that we will consider
the whole parameter space of the MSSM
with the only assumption of universality,
i.e. $m$, $M$, $A$, $B$. Let us remark, however, that the constraints
reviewed above are general and they could also be applied for the 
non--universal case. In Fig.1 we have presented in detail, as a guiding
example, the 
(well--known minimal supergravity) case $B=A-m$ with $m$=100 GeV to get an idea
of how strong the different constraints are and then we will vary $B$ and $m$
freely
in order to obtain the most general results. 
The excluded regions are plotted in the remaining parameter space
($A/m$, $M/m$). 
It is worth 
noticing here that even before imposing CCB and UFB constraints, the parameter
space is strongly restricted by the experiment
as explains in the Figure caption.
The restrictions coming from the UFB constraints 
are very strong. {\it Most} of the parameter space is in fact excluded by 
the UFB--3 constraint. 
Notice from Fig.1 that there are areas that are simultaneously constrained
by different types of bounds. 
Besides, the values of $A$ and $M$ are both bounded from below and above
in a correlated way.
At the end of the day, the allowed region left
(white) is quite small.


\begin{figure}[htb]
\centerline{
\epsfig{file=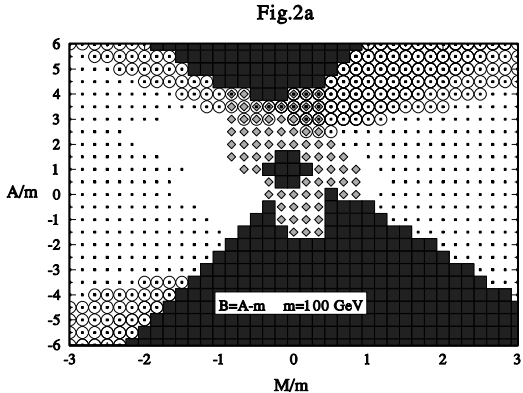,height=11.5cm,angle=180
}}
\vspace{-3.5cm}
\caption{
Excluded regions in the parameter space of the MSSM,
with $M^{\rm phys}_{\rm top}=174$ GeV.
The central black region is excluded because there is no solution
for $\mu$ capable of producing the correct electroweak breaking.
The upper and lower black regions are excluded because it is not possible
to reproduce the experimental mass of the top due to the infrared fixed
point of $\lambda_{top}$.
The filled diamonds
indicate regions excluded by the experimental lower bounds on SUSY particle
masses.
The small filled squares indicate regions excluded by 
UFB constraints,
mainly the UFB-3 one.
The circles indicate regions excluded
by CCB constraints.} 
\end{figure}


In order to show now that the CCB and UFB constraints put important
bounds not only on the value of $A$ and $M$, but also on the values
of $B$ and $m$, we generalize
the previous analysis by varying first the value of $B$. For
a particular value of $m$, the larger the value of $B$ the smaller the
allowed region becomes. In general, for      
$m\simlt 500$ GeV 
(larger values of $m$ would conflict absence--of--fine--tuning
requirements for electroweak breaking),
$B$ has to satisfy the bound
$|B|\simlt 3.5\ m$.
%
%
The results also indicate
that the smaller the value of $m$, the more restrictive the constraints
become. 
In fact, it is possible to find a value of $m$ for which the whole
parameter space turns out to be excluded. This interesting lower bound
on $m$ is 
$m\geq 50\ GeV$.  
%
%
{}From this discussion it is evident that the limiting case $m=0$
is also excluded. Of course, this has obvious implications for no--scale
supergravity models since that limit is usually obtained. 
Figures illustrating these numerical results, as well as a discussion about 
the physical reasons underlying them, can be found in \cite{CCB1}.

Finally, let us remark that the previous analyses were performed assuming
universality of the soft terms at $M_{GUT}$. 
As mentioned in the introduction, 
the larger the initial scale for the running of the soft terms
is, 
the stronger the bounds become.
This can be understood from our discussion about the UFB--3
direction above: the larger the initial scale for the running is, the
more important the negative contribution $m_2^2-{\mu}^2$ to the potential
(see (\ref{SU8}) and (\ref{SU9})) becomes.
In particular,
in the standard supergravity framework, where SUSY is broken in 
a hidden sector, the natural initial scale to implement the boundary
conditions for the soft terms is $M_P\equiv M_{Planck}/\sqrt{8\pi}$
rather than $M_{GUT}$. Using the scale $M_P$ the constraints are substantially
increased. For instance, regions of large $M$ which were previously allowed
for $m>100$ GeV
become now completely excluded, also 
the above bounds on $B$ and $m$ 
become $|B|\simlt 3m$ and $m\geq 55$ GeV respectively.
Figures illustrating these results can be found in \cite{CCB2}. 

The CCB and UFB constraints can be applied to particular SUSY scenarios
as mentioned in the introduction.
For instance, 
in the case of the infrared fixed point model, 
the parameter space turns out to be severely constrained, including
the bound $|M/m|\simlt 1$. Figures can
be found again in \cite{CCB2}. 
SUSY/string scenarios are reviewed in the next section.

\section{Charge and color breaking in superstrings}

Let us briefly review the basic ingredients required for this analysis.
First we will concentrate on the form of soft SUSY--breaking terms.
The
general form of the soft SUSY--breaking Lagrangian in the context of the
MSSM for instance is given by
${\cal L}_{soft}=\frac{1}{2}(\sum_a M_a 
{\lambda}_a {\lambda}_a + h.c.)
- V_{\rm soft}$,
%
%
where 
$\lambda_a$ are gaugino canonically normalized fields
and $V_{\rm soft}$ is given in (\ref{Vsoft}).
The above soft parameters are free in the context of the pure
MSSM but 
can be obtained dynamically in a supergravity theory through
the spontaneous breaking of local SUSY in a hidden sector \cite{BIM3}.
In supergravity models 
obtained from superstring compactifications
there is a natural hidden sector built--in: the complex dilaton field
$S$ and the complex moduli fields $T_i$. 
Assuming that the auxiliary fields of those multiplets are the seed
of SUSY breaking, interesting predictions about soft terms are obtained.
Let us first focus on the very interesting 
case where the dilaton field is
the source of all the SUSY breaking 
\cite{BIM3}. 
Since, at string tree--level, 
the dilaton couples in a universal manner to all particles, 
this limit is quite model {\it independent}. 
The soft parameters
are:
%
%
%
%
$m_{\alpha}^2 = m_{3/2}^2$, $M_a = \pm \sqrt{3} \; m_{3/2}$,
$A_{\alpha\beta\gamma} = - M_a$,
where $A_{\alpha\beta\gamma}=A_u, A_d, A_e$ in a self--explanatory notation.
This dilaton--dominated scenario 
is attractive for its simplicity and for
the natural explanation that it offers to the {\it universality} of the
soft terms. 
Since the value of $B$ 
is more model dependent, 
it is better to take it as a free parameter in order to carry out the most
general analysis.

The second basic ingredient of our analysis concerns the constraints
associated with the existence of dangerous directions in the field
space. These were explained in section 2 for a generic
SUSY theory and therefore can be applied for any four--dimensional
superstring model.
In the particular case of the dilaton--dominated scenario,
the restrictions coming from the UFB constraints are very strong and
the
whole parameter space ($m_{3/2}$,$B$) turns out to be excluded. Most of it is in fact 
excluded by the UFB--3 constraint. Figures illustrating this result
can be found in \cite{CCB3}.


Given the above dramatic conclusions about the dilaton--dominated
scenario, let us briefly discuss a possible way--out. The most
straightforward possibility is to assume that also the moduli
fields $T_i$ contribute to SUSY breaking, which is in fact a more general
situation. Then
the soft terms are modified, new free parameters beyond 
$m_{3/2}$ and $B$ appear, and possibly some regions in the
parameter space will be allowed. 
This situation is more model dependent since different compactification
schemes have different numbers and types of moduli and different couplings
of them to matter, therefore giving rise to different soft terms.
In the simple case of (0,2) symmetric Abelian orbifolds with diagonal
moduli and matter metrics the soft terms have been computed.
To assume that SUSY breaking is equally shared among $T_{i}$'s, i.e. 
the ``overall modulus'' $T$ scenario is a good starting point in the
analysis of charge and color breaking since essentially only one more
free parameter must be added \cite{prepa}.

\section{Final comments and outlook}

We have shown in this review that charge and color breaking constraints
on the parameter space of generic SUSY theories are very strong. This is
particularly true in the case of SUSY theories deriving from weakly 
coupled heterotic
superstring where information about the structure of soft terms can be
obtained. Since the dilaton--dominated SUSY--breaking scenario is excluded
on these grounds, it would be very interesting
to study possible way--outs to the previous dramatic conclusion.
As mentioned in section 3 one possibility is to asume that also the moduli
fields contribute to SUSY breaking \cite{prepa}. 
Another possibility, 
is 
to think that the perturbative and non-perturbative corrections
to the ``standard'' string tree--level dilaton--dominated scenario 
are important
and can modify the previous conclusions. 
Actually, some
one--loop string corrections 
have been
calculated for orbifold models 
and they are rather small
for sensible values of the moduli. 
However,
this is not the case for the string non-perturbative corrections, whose
size could be much larger.  
modifying in a sensible way the formulas for soft terms
\cite{dilatonio}. 
Finally, recently some information has been obtained 
in the sense that all superstring theories
seem to correspond to some points in the parameter space of a unique
strongly coupled eleven--dimensional underlying theory, M--theory.
Once the structure of soft terms is known
charge and color breaking constraints should be applied to determine
their phenomenological viability.

All the strongs constraints on the soft parameter space of SUSY
theories that have been reviewed here come from the requirement that
the standard vacuum is the global minimum of the theory.
In this sense, one possibility to avoid some of the above constraints
is to accept that we live in a metastable vacuum, provided that its
lifetime is longer than the present age of the universe \cite{claudson}, thus
rescuing points in the parameter space. 
In order to carry out this study one might consider two possibilities:
quantum tunneling at zero temperature from the standard vacuum to the
charge and color breaking one and thermal effects in the hot early universe.
Regarding the latter, although 
there is
a thermal energy to cross the barrier, due to the high temperature of the early
universe, the barrier is also higher \cite{riotto}.
In the case of quantum tunneling at zero temperature,
for instance the CCB minima
associated with the top--quark Yukawa coupling are the only ones to which
the standard vacuum might decay within the lifetime of the 
universe \cite{claudson,riotto,langacker,strumia}.
The CCB minima associated with other Yukawas are deeper 
but the
height of the barrier ($h\sim 1/\lambda^2$) 
is too large to allow an efficient tunnelling
probability ($\sim e^{-ch}$). 
In this sense the bounds that we reviewed here are basically 
the most conservative ones (in the sense of safe ones). Needless to say
that in any case, the identification of the dangerous CCB and UFB minima
is the first necessary step for the cosmological analysis.
In the context of gauge--mediated SUSY--breaking models the
standard vacuum seems to be stable cosmologically, but only if 
certain couplings are sufficiently small \cite{randall}.

Let us remark however that the possibility of living in a metastable vacuum
poses several problems. First of all, as was first suggested in \cite{CCB3},
it is hard to understand 
how  the cosmological constant is vanishing precisely
in such local ``interim'' vacuum.  
Even if a solution to that problem is found we would still
have to face the rather bizarre (but mathematically possible) 
situation of a future cosmological catastrophe, which does not seem
very attractive (at least for our descendants!).
Finally, from a more scientific (and less philosophical) point of view 
one needs to explain (without invoking an
anthropic principle) how does the universe manage to reach the 
standard minimum in the first place in spite of being local and metastable.
This requires the analysis of all possible cosmological scenarios.
In particular one can consider scenarios where the initial conditions
are dictated by thermal effects or inflationary scenarios. In the former
the standard vacuum is the closest one to the origin and therefore it
is the thermal equilibrium state at large temperatures \cite{claudson}. 
The inflationary
scenario may be much more dangerous and involved due to large fluctuations
of all the scalar fields, that could be driven in this way to the
dangerous minima. Whether this is the case or not is a complex issue
that is hotly discussed \cite{vilja}.

\section*{Acknowledgments}

Research supported in part by KOSEF, under the Brainpool Program;
the CICYT, under contract
AEN93-0673; the European Union,
under contracts CHRX-CT93-0132 and
SC1-CT92-0792.

\def\MPL #1 #2 #3 {{\em Mod.~Phys.~Lett.}~{\bf#1}\ (#2) #3 }
\def\NPB #1 #2 #3 {{\em Nucl.~Phys.}~{\bf B#1}\ (#2) #3 }
\def\PLB #1 #2 #3 {{\em Phys.~Lett.}~{\bf B#1}\ (#2) #3 }
\def\PR  #1 #2 #3 {{\em Phys.~Rep.}~{\bf#1}\ (#2) #3 }
\def\PRD #1 #2 #3 {{\em Phys.~Rev.}~{\bf D#1}\ (#2) #3 }
\def\PRL #1 #2 #3 {{\em Phys.~Rev.~Lett.}~{\bf#1}\ (#2) #3 }
\def\PTP #1 #2 #3 {{\em Prog.~Theor.~Phys.}~{\bf#1}\ (#2) #3 }
\def\RMP #1 #2 #3 {{\em Rev.~Mod.~Phys.}~{\bf#1}\ (#2) #3 }
\def\ZPC #1 #2 #3 {{\em Z.~Phys.}~{\bf C#1}\ (#2) #3 }


\begin{thebibliography}{99}
%
\bibitem{reciente}For a recent review, see: J.A. Casas, 
{\it hep-ph/9707475}, and references therein
%
\bibitem{CCB1} J.A. Casas, A. Lleyda and C. Mu\~noz, 
\NPB 471 1996 3
%
\bibitem{olive} T. Falk, K. Olive, L. Roszkowski and M. Srednicki,
\PLB367 1996 183
%
\bibitem{baer} H. Baer, M. Brhlik and D. Casta\~no,
\PRD 54 1996 6944
%
\bibitem{CCB2} J.A. Casas, A. Lleyda and C. Mu\~noz, 
\PLB 389 1996 305
%
\bibitem{strumia} A. Strumia, \NPB 482 1996 24
%
\bibitem{randall} I. Dasgupta, B.A. Dobrescu and L. Randall, 
\NPB 483 1997 95 
%
\bibitem{dimo} J.A. Casas and S. Dimopoulos, \PLB 387 1996 107
%
\bibitem{BIM3}For a recent review, see: 
A. Brignole, L.E. Ib\'{a}\~{n}ez and C. Mu\~noz,
{\it hep-ph/9707209}, and references therein
%
%
\bibitem{CCB3} J.A. Casas, A. Lleyda and C. Mu\~noz, 
\PLB 380 1996 59
%
%
%
\bibitem{prepa} A. Ibarra, J.A. Casas and C. Mu\~noz, in preparation
%
\bibitem{dilatonio} J.A. Casas, \PLB 384 1996 103
%
\bibitem{claudson}M. Claudson, L.J. Hall and I. Hinchliffe, \NPB 228 1983 501
%
\bibitem{riotto} A. Riotto and E. Roulet, \PLB 377 1996 60
%
\bibitem{langacker} A. Kusenko, P. Langacker and G. Segre, \PRD 54 1996 5284;
A. Kusenko and P. Langacker, \PLB 391 1997 29
%
\bibitem{vilja} A. Riotto, E. Roulet and I. Vilja,
\PLB 390 1997 73; T. Falk, K. Olive, L. Roszkowski, A. Singh and M. Srednicki, 
\PLB 396 1997 50
%



\end{thebibliography}
\end{document}